\pdfoutput=1

\documentclass[aip,reprint,graphicx]{revtex4-1}

\usepackage{graphicx}

\usepackage[utf8]{inputenc}

\usepackage{SIunits}

\draft

\begin{document}

\title{Closed-cycle, low-vibration 4\,K cryostat for ion traps and other applications}

\author{P.~Micke}
\email[Author to whom correspondence should be addressed: ]{peter.micke@mpi-hd.mpg.de}

\affiliation{Max-Planck-Institut f\"ur Kernphysik, Saupfercheckweg 1, 69117 Heidelberg, Germany}
\affiliation{Physikalisch-Technische Bundesanstalt, Bundesallee 100, 38116 Braunschweig, Germany}

\author{J.~Stark}
\affiliation{Max-Planck-Institut f\"ur Kernphysik, Saupfercheckweg 1, 69117 Heidelberg, Germany}
\affiliation{Heidelberg Graduate School of Fundamental Physics, Ruprecht-Karls-Universität Heidelberg, Im Neuenheimer Feld 226, 69120 Heidelberg, Germany}

\author{S.~A.~King}
\affiliation{Physikalisch-Technische Bundesanstalt, Bundesallee 100, 38116 Braunschweig, Germany}

\author{T.~Leopold}
\affiliation{Physikalisch-Technische Bundesanstalt, Bundesallee 100, 38116 Braunschweig, Germany}

\author{T.~Pfeifer}
\affiliation{Max-Planck-Institut f\"ur Kernphysik, Saupfercheckweg 1, 69117 Heidelberg, Germany}

\author{L.~Schm\"oger}
\affiliation{Max-Planck-Institut f\"ur Kernphysik, Saupfercheckweg 1, 69117 Heidelberg, Germany}
\affiliation{Physikalisch-Technische Bundesanstalt, Bundesallee 100, 38116 Braunschweig, Germany}

\author{M.~Schwarz}
\affiliation{Max-Planck-Institut f\"ur Kernphysik, Saupfercheckweg 1, 69117 Heidelberg, Germany}
\affiliation{Physikalisch-Technische Bundesanstalt, Bundesallee 100, 38116 Braunschweig, Germany}

\author{L.~J.~Spie{\ss}}
\affiliation{Max-Planck-Institut f\"ur Kernphysik, Saupfercheckweg 1, 69117 Heidelberg, Germany}

\author{P.~O.~Schmidt}
\affiliation{Physikalisch-Technische Bundesanstalt, Bundesallee 100, 38116 Braunschweig, Germany}
\affiliation{Institut für Quantenoptik, Leibniz Universität Hannover, Welfengarten 1, 30167 Hannover, Germany}

\author{J.~R.~{Crespo L\'opez-Urrutia}}
\affiliation{Max-Planck-Institut f\"ur Kernphysik, Saupfercheckweg 1, 69117 Heidelberg, Germany}

\date{\today}

\begin{abstract}
\emph{In-vacuo} cryogenic environments are ideal for applications requiring both low temperatures and extremely low particle densities. This enables reaching long storage and coherence times for example in ion traps, essential requirements for experiments with highly charged ions, quantum computation, and optical clocks. We have developed a novel cryostat continuously refrigerated with a pulse-tube cryocooler and providing the lowest vibration level reported for such a closed-cycle system with 1\,W cooling power for a $< 5$\,K experiment. A decoupling system suppresses vibrations from the cryocooler by three orders of magnitude down to a level of 10\,nm peak amplitudes in the horizontal plane. Heat loads of about 40\,W (at 45 K) and 1W (at 4 K) are transferred from an experimental chamber, mounted on an optical table, to the cryocooler through a vacuum-insulated massive 120 kg inertial copper pendulum. The 1.4\,m long pendulum allows installation of the cryocooler in a separate, acoustically isolated machine room. In the laser laboratory, we measured the residual vibrations using an interferometric setup. The positioning of the 4\,K elements is reproduced to better than a few \micro m after a full thermal cycle to room temperature. Extreme high vacuum on the $10^{-15}$\,mbar level is achieved. In collaboration with the Max-Planck-Intitut für Kernphysik (MPIK), such a setup is now in operation at the Physikalisch-Technische Bundesanstalt (PTB) for a next-generation optical clock experiment using highly charged ions.
\end{abstract}

\pacs{}

\maketitle

\section{Introduction}\label{sec:I}
Cryogenic temperatures are needed in many fields of research, such as in studies of superconductivity and superfluidity or the development of low-noise detectors and electronics. They can also generate extreme high vacuum (XHV) conditions, an advantage exploited in cryogenic Penning traps \cite{repp_pentatrap:_2012,sturm_high-precision_2017} or Paul traps \cite{schwarz_cryogenic_2012,niedermayr_cryogenic_2014,pagano_cryogenic_2019}, and nearly impurity-free vacua in multipole radio-frequency (rf) traps \cite{gerlich_ion-neutral_1995}. Furthermore, superconducting detectors for fluorescence photons integrated into surface Paul traps for scalable quantum information processing require operation at cryogenic temperatures \cite{slichter_uv-sensitive_2017,wollman_uv_2017}. For these reasons cryogenics is gaining ground in the last years within the atomic- and quantum-physics community. Moreover, an extremely low residual gas density is crucial for suppressing charge-exchange-induced ion recombination\cite{muller_scaling_1977} of highly charged ions (HCIs), a key aspect for the current development of optical atomic clocks using HCIs \cite{kozlov_highly_2018}. Recently, re-trapping of HCIs in a cryogenic Paul-trap experiment (CryPTEx)\cite{schwarz_cryogenic_2012,schmoger_coulomb_2015,schmoger_deceleration_2015} and sympathetic cooling of those HCIs through interactions with a laser-cooled Coulomb crystal were demonstrated at the Max-Planck-Institut f\"ur Kernphysik (MPIK). High-resolution spectroscopy of HCIs using advanced spectroscopic methods like quantum logic\cite{schmidt_spectroscopy_2005} requires that collisions have to be avoided for many minutes. This is also important for quantum-physics research, e.\,g., in scalable setups for quantum simulations or quantum computing\cite{kim_system_2005,niedermayr_cryogenic_2014,alonso_generation_2016,brown_co-designing_2016,brandl_cryogenic_2016-1,pagano_cryogenic_2019}, since collisions with background gas molecules or atoms cause decoherence or even unacceptable ion losses, which are suppressed at cryogenic temperatures. In addition, a cryogenic trap suppresses ion heating\cite{deslauriers_scaling_2006,labaziewicz_suppression_2008,brownnutt_ion-trap_2015} due to reduced electric field noise. Optical atomic clocks aiming at sub-$10^{-18}$ fractional frequency inaccuracy also benefited in recent years from the strong suppression of residual systematics in a cryogenic environment, such as the ac Stark shift from black-body radiation and collisional shifts due to background gas\cite{rosenband_frequency_2008,middelmann_tackling_2011,gibble_scattering_2013,ushijima_cryogenic_2015,takamoto_frequency_2015,ludlow_optical_2015,vutha_collisional_2017}.

Inside a closed vessel the total vapor pressure given by the sum of partial pressures is tremendously reduced by cooling down the entire vessel or some surfaces therein. At temperatures below the boiling point of nitrogen (77\,K at normal pressure) only neon, helium, and hydrogen are still gases. At 4\,K, all gases are either condensed, frozen out or cryo-adsorbed. Residual gas particles impinging on cryogenic surfaces lose kinetic energy, stick to the surface (dubbed cryopumping) and do not outgas (desorb). At around 4\,K, XHV pressures below $10^{-14}$\,mbar are routinely achieved\cite{benvenuti_characteristics_1974}. Values down to the $10^{-17}$\,mbar level were reported\cite{gabrielse_thousandfold_1990,diederich_observing_1998} for sealed cryogenic traps. 

Two methods are chiefly used to reach 4\,K temperatures. The first is liquid helium (LHe) refrigeration in continuous-flow or bath cryostats (see, e.\,g., \cite{ekin_experimental_2006} and the references therein; Refs. \cite{brandl_cryogenic_2016-1} and \cite{poitzsch_cryogenic_1996} as sample applications). For pre-cooling and also for the outer heat shields of such systems, liquid nitrogen (LN$_2$) is commonly used. The inner stage is then cooled with LHe down to 4\,K. There is the option of achieving lower temperatures through so-called lambda-point refrigeration under pumping (down to approximately 1.8\,K \cite{ekin_experimental_2006}), subsequent cooling through adiabatic demagnetization or in $^3$He-$^4$He mixing cryostats. Inconveniently, since LHe and LN$_2$ continuously evaporate from these open systems, regular refilling is needed. Evaporation of LHe at the rate of approximately one liter every hour per W of cooling power at 4\,K causes considerable running costs. The LHe and LN$_2$ reservoirs, their thermal shields and insulation are bulky and hinder access to the experimental region. In many cases, the cryogenic support of the cold mass and reservoirs is feeble in order to reduce the thermal load from room temperature. This can cause sensitivity to vibrations and mechanical shifts due to thermal expansion and the changing level of the cryogen.   

As a second method, continuously operating closed-cycle refrigerators supplied by various vendors can be employed (see, e.\,g., \cite{radenbaugh_refrigeration_2004,ekin_experimental_2006,radebaugh_cryocoolers:_2009} and references therein). They do not need refilling and offer very stable temperatures during operation. Frequently used types are Gifford-McMahon and pulse-tube cryocoolers, both typically as two-stage systems providing cooling powers on the order of 40\,W at the first stage (around 40\,K) and 1\,W at the second stage (around 4\,K). A major drawback of such systems is the cycling flow (around 1\,s period) of strongly pressurized cooling gas, resulting in mechanical vibrations and noise at the cold head and other parts. Many cryogenic experiments are sensitive to mechanical vibrations on a level of sub-\micro m: laser spectroscopy of trapped atoms or ions \cite{middelmann_tackling_2011,brandl_cryogenic_2016-1,pagano_cryogenic_2019}, silicon cavities for optical oscillators\cite{kessler_sub-40-mhz-linewidth_2012,matei_1.5_2017,robinson_crystalline_2018}, sapphire oscillators\cite{grop_10_2010,hartnett_ultra-low-phase-noise_2012}, frequency references based on spectral hole burning \cite{thorpe_frequency_2011,chen_spectrally_2011,cook_laser-frequency_2015}, x-ray monochromators\cite{mochizuki_cryogenic_2001,toellner_cryogenically_2006},  bolometers and detectors searching for, e.\,g., neutrinoless double-beta decay \cite{cuore_collaboration_first_2018} or dark matter \cite{aguirre_dark_2015}, instruments and microcalorimeters for infrared, x-ray and gamma astronomy\cite{radebaugh_cryocoolers:_2009}, and gravitational wave detectors\cite{somiya_detector_2012}.  
For such applications, pulse-tube cryocoolers\cite{radebaugh_pulse_2003} offer lesser inherent vibrations than Gifford-McMahon cryocoolers and are more reliable since there are no movable parts at the cold head itself\cite{ekin_experimental_2006,radebaugh_cryocoolers:_2009}. Typical maintenance intervals are on the order of 20\,000 hours for the cold head and 30\,000 hours for its compressor. Since a 1\,Hz alternating compression and expansion of He gas driven by a rotary valve is needed for the cooling, vibrations at the cold head with amplitudes at the 10\,\micro m level are still induced.

Commercial vibration-insulation systems are available for closed-cycle refrigerators. Some use a He gas heat-exchange unit, i.\,e. a He volume enclosed by edge-welded or rubber bellows, as thermal link between the cryocooler and the cold stage (see, e.\,g., \cite{sage_loading_2012,olivieri_vibrations_2017,pagano_cryogenic_2019}). Except for the flexible bellows, no rigid mechanical contact exists. The He gas needs continuous replenishment to maintain a pressure of typically 30\,mbar above atmospheric pressure. It is crucial to operate such a gas-exchange unit above the He boiling point of 4.2\,K to prevent mechanical contact through condensed He \cite{dubielzig_notitle_nodate}. Consequently, the experiment has to be operated at a sufficiently high temperature, in cases several kelvins above 4.2\,K\cite{pagano_cryogenic_2019} depending on the thermal coupling of the experiment to the gas-exchange unit. This comes at the price of a reduced cryopumping performance. Some research groups have developed vibration-decoupling systems. Active noise cancellation of pulse-tube cryocooler vibrations was implemented \cite{daddabbo_active_2018}. However, most groups have applied passive decoupling schemes based on flexible metallic links\cite{ikushima_ultra-low-vibration_2008,schwarz_cryogenic_2012,dhuley_thermal_2017} which must satisfy the conflicting needs of high thermal but low mechanical coupling.

All this considered, we started developing a cryogenic system with vibration decoupling based on our experience from CryPTEx\cite{schwarz_cryogenic_2012} with the goal of improving mechanical stability and lowering the achievable temperatures. A pulse-tube cryocooler ensures reliable, low-maintenance operation over months, low running costs, and stable ion trap temperatures. Our design achieves high reproducibility of the trap position after thermal cycling. By installing the cryocooler and other noisy components in a separate room, the Paul trap mounted onto a pneumatically floating optical table in the adjacent laser laboratory meets the stability demands for optical atomic clocks. It is already in operation for studying forbidden optical transitions of HCIs. These are produced by a Heidelberg Compact Electron Beam Ion Trap (HC-EBIT) \cite{micke_heidelberg_2018}, transferred through an ion beamline, decelerated, and re-trapped in a cryogenic Paul trap \cite{leopold_cryogenic_nodate}.

In this paper we describe the cryogenic system in detail which we initially designed and assembled at MPIK in Heidelberg, Germany. After completion, we moved it to the Physikalisch-Technische Bundesanstalt (PTB) in Braunschweig, Germany, where various measurements reported here characterized its performance regarding temperatures, vacuum level and residual vibrations. Currently, the setup is operated there at the QUEST Institute for a cryogenic linear Paul trap, aiming at quantum logic spectroscopy of HCIs. A similar setup was also recently assembled by our collaborators at Aarhus university for the spectroscopy of molecules in a cryogenic Paul trap. Another slightly modified version is currently under construction at MPIK. Its intended application is HCI trapping in a monolithic radio-frequency trap and resonator\cite{stark_et_al._superconducting_nodate}, aiming at direct frequency-comb spectroscopy in the extreme ultra-violet range using high-harmonic generation in an optical enhancement cavity\cite{nauta_towards_2017}.

\section{Design}\label{sec:CD}

\subsection{Concept}

\begin{figure*}
	\includegraphics[width=0.95\linewidth]{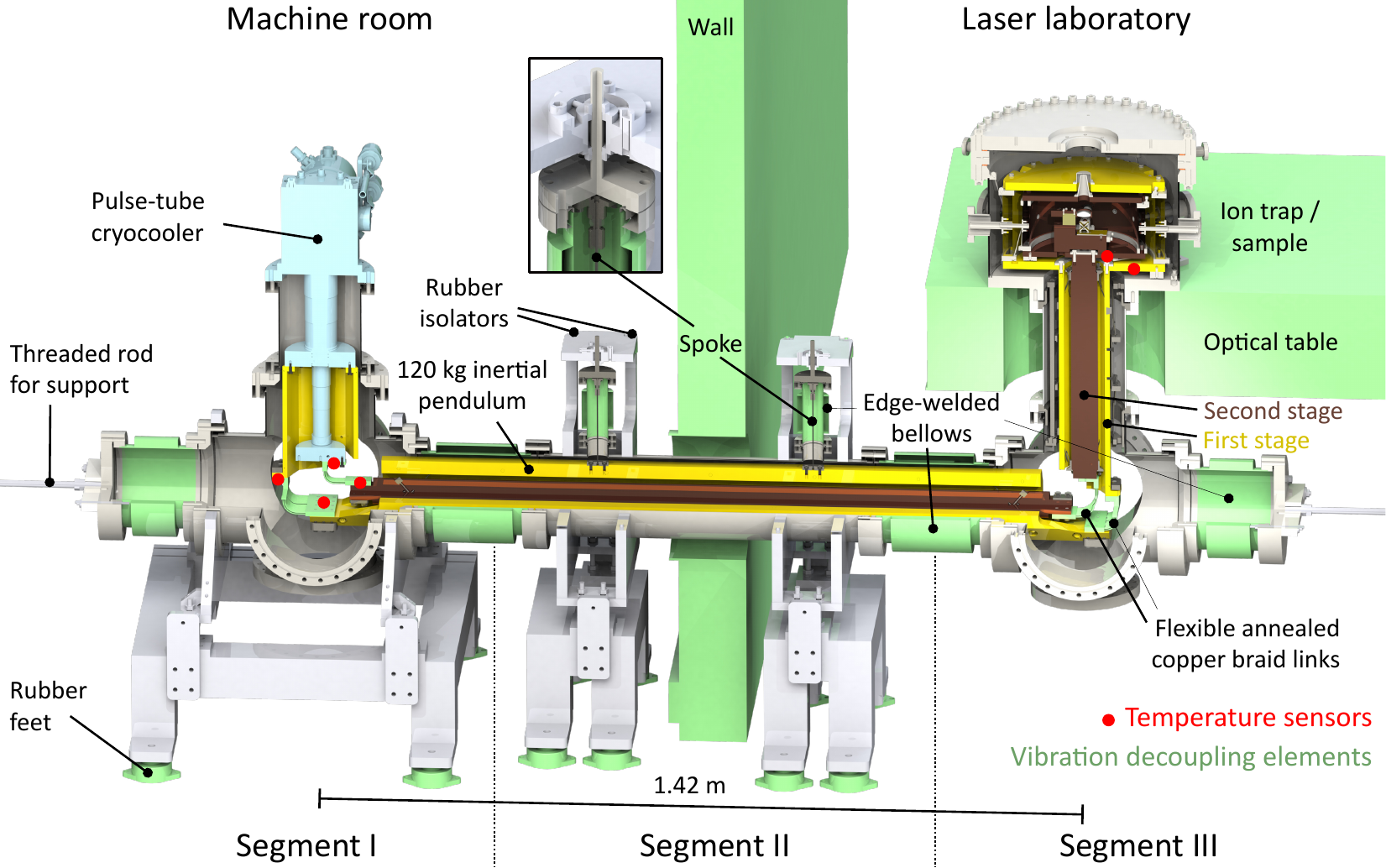}
	\caption{Cutaway drawing of the cryogenic system\label{fig:overview}. The system extends over two rooms separated by a noise-insulating wall, with the pulse-tube cryocooler located in the machine room on the left and the cooled ion trap\,/\,sample on the right on top of a pneumatically floating optical table in the laser laboratory. The cryogenic system is subdivided into three segments. The heat shields which enclose the second stage vibration-decoupling sections with the flexible copper links are not shown for visibility. The inset magnifies the position adjustment of the inertial pendulum. See text for further details.}%
\end{figure*}

Detrimental vibrations can propagate from the cold head and pumping system through air, the floor, the vacuum chambers and the thermal transfer elements. We address these paths in our design by (a) acoustically separating the noisy elements from the quiet area, (b) installing the various components on vibration-reducing supports attached to the floor, (c) introducing highly flexible bellows in the vacuum system and (d) using a low-pass, heavy inertial vibration filter as the thermal transfer unit. In the following we will describe these measures.

The cryogenic system (an overview is shown in Fig.\,\ref{fig:overview}) extends over two rooms which are separated by an acoustically insulating wall. One of them is the laser laboratory, in which the experimental chamber, containing a linear Paul trap \cite{leopold_cryogenic_nodate}, is rigidly mounted on top of a pneumatically floating optical table to decouple it from the floor. Although this cryogenic system was intended for use with an ion trap, any object of study could be cooled with it. For simplicity, we will refer to this object as just the \textit{ion trap} in this article. In the adjacent room, referred to as the \textit{machine room}, a two-stage closed-cycle pulse-tube cryocooler is installed at a distance of 1.4\,m from the ion trap region, which is larger than in most other cryogenic setups. The helium compressor is located at the far end of the machine room for better vibration suppression. A vacuum tube containing the thermal transfer unit (TTU) runs through the wall and connects the two sections of the cryogenic system. Massive high-purity oxygen-free high thermal conductivity (OFHC) copper parts, linked by two flexible vibration-decoupling sections, are used in the TTU between the cryocooler and the ion trap. Together, they constitute an inertial low-pass vibration filter. Moreover, the rigidity of the vacuum chamber surrounding the TTU is also broken at several locations by introducing flexible edge-welded bellows that absorb the noise components above their lowest oscillation frequencies (a few Hz). Hence, several vacuum chambers ``float'' between bellows, with their mass also acting as low-pass filter for vibrations. Due to our space and technical constraints, such as the required orientation of the pulse tube, the system is divided into three segments: I (vertical) with the cryocooler, II (horizontal) with the inertial pendulum, and III (vertical) with the TTU connecting to the ion trap. The cold masses are divided in two stages: the first one (35 to 50\,K) is the thermal heat shield surrounding the second one (3.5 to 5.0\,K), which cools the ion trap. The thermal insulation vacuum is differentially connected with the vacuum of the ion trap where an XHV level is required once the system is cold. All mechanical pumps for initial pumping and maintaining an ultra-high vacuum (UHV) room-temperature base pressure are located in the machine room. The vacuum chambers enclosing the TTU are vibration-decoupled by six edge-welded bellows in total. Since the diameter of the larger DN160CF bellows is around 170\,mm, strong forces arise when the vacuum system is pumped out due to the outer atmospheric pressure. Two outer counteracting DN160CF bellows in extension of the segment II are attached on their outer flanges to rigid pillars anchored to the ground of each room to balance the horizontal pressure forces, leaving zero net forces acting on the vacuum chambers. The chamber of segment I is mounted on an aluminum frame supported by damping rubber feet to the noise-damping screed used in the floor of the machine room. Segment II runs horizontally at 50\,cm height through a hole in the noise-insulating wall and is also held by two aluminum frames, one in the machine room and the second in the laser laboratory. Both are again supported by damping rubber feet. The vacuum chamber of segment III runs from below through a hole in the aforementioned floating optical table and is stiffly mounted to its top, the only rigid connection of this chamber. This arrangement provides 360 degree optical access on the table top in the horizontal plane and leaves also the top lid of the ion-trap chamber free, offering convenient access to the inside. 

\begin{figure*}
	\includegraphics[width=0.95\linewidth]{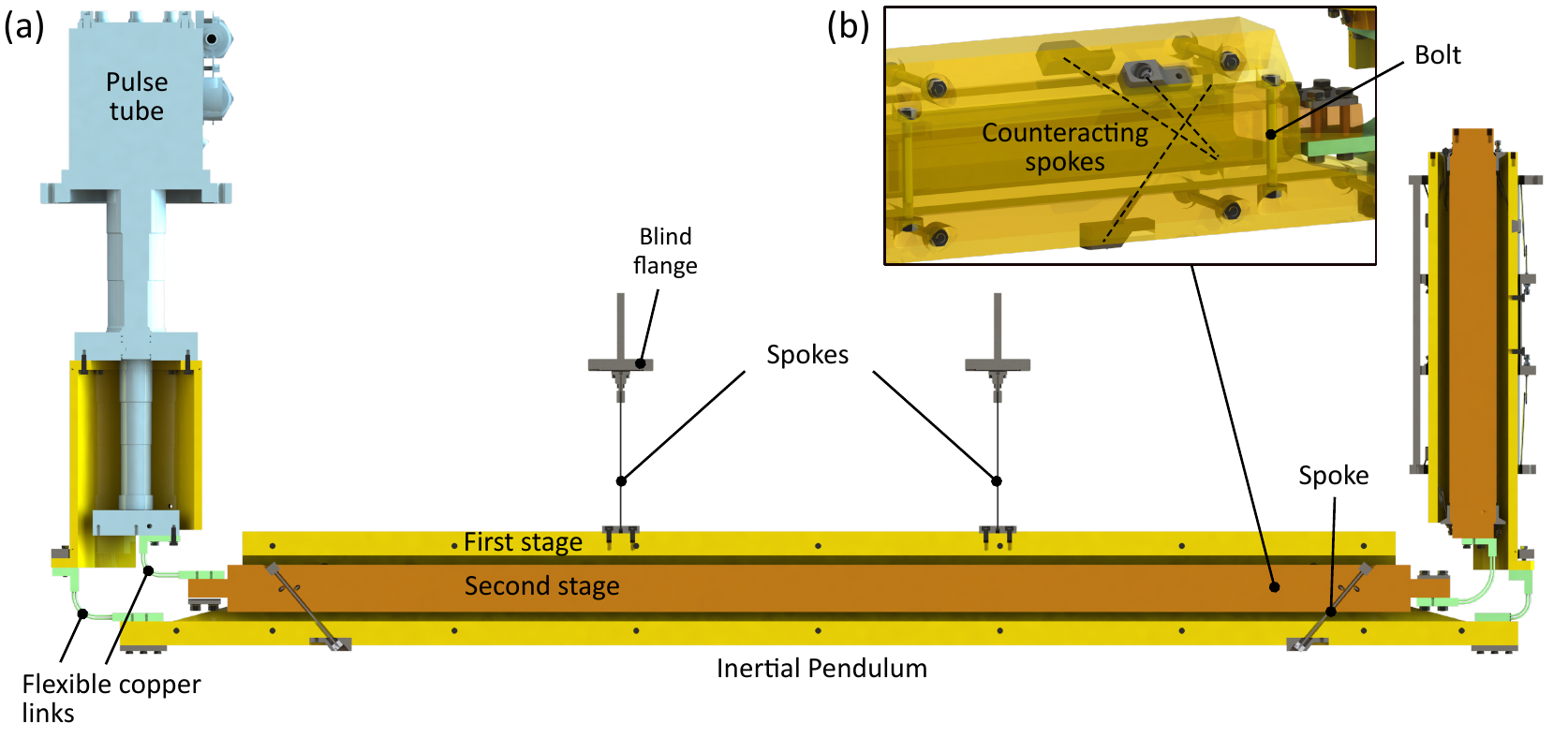}
	\caption{(a) Cross section of the thermal transfer unit (TTU) \label{fig:TL} with its two temperature stages. The inner parts of the three segments are connected by flexible copper links. Interlaced heat shields enclosing the vibration-decoupling sections are not shown. (b) Mounting structure of the inertial pendulum (both stages are depicted as transparent for visibility). The hollow first stage of the pendulum is assembled from prismatic copper bars bolted together. As second stage, a massive copper rod is held inside the first stage by six counteracting spokes. See text for further details.}%
\end{figure*}

\subsection{Vacuum system}\label{sec:PS}
In total, the system comprises three vacuum chambers: two custom-made six-way crosses with DN200CF ports and the DN160CF tube of segment~II with two vertical DN40CF flanges. Additionally, two DN160CF tubes are used in segments~I and III, respectively (see Fig.\,\ref{fig:overview}). All parts are made of low-magnetic stainless steel and are electro-polished to reduce outgassing and the radiation-emission coefficient, thus reducing the radiative heat transfer to the cryogenic elements. In segment~I, the pulse-tube cryocooler is mounted on a DN160CF vertical tube attached to the top of the first six-way cross, which is connected to segment II by means of an edge-welded DN160CF bellows with 40 diaphragm pairs. On the opposite side of segment~II, an identical one connects to the second six-way cross (of segment~III). Attached to the top flange of this cross, a second vertical DN160CF tube holds the in-vacuum room-temperature frame of the last segment of the TTU (detailed in Sec. \ref{sec:TL_SIII}). This tube passes through the 350-mm diameter hole in the optical table and is connected to the bottom of the ion-trap chamber \cite{leopold_cryogenic_nodate}. For balancing the compressive forces, two edge-welded DN160CF bellows with 20 diaphragm pairs are mounted to the outer flanges of each of the two six-way crosses. These bellows are terminated with blind flanges fixed by means of M12-threaded rods to massive pillars anchored to the floor in each of the two rooms. Two edge-welded DN40CF bellows are installed vertically on the horizontal chamber, with their top flanges closed by modified DN40CF blind flanges. On their vacuum side, each one holds a 150-mm long spoke from which the inertial pendulum of the TTU is suspended (see Sec. \ref{sec:TL_SII} for further details). On the air side, adjustable threaded rods are welded. They run vertically through oversize holes on aluminum plates, with nuts and larger 25\,mm-diameter washers allowing for horizontal position adjustment of the TTU by means of three set screws (see inset of Fig. \ref{fig:overview}). The height is adjusted by turning the nut on the threaded rod, lifting or lowering the pendulum.

A single 300\,l/s turbomolecular pump (TMP) attached to segment~I, backed by a smaller 70\,l/s TMP and an oil-free scroll pump is used for pumping. This two-stage TMP system increases the H$_2$ compression ratio by more than three orders of magnitude compared to a single-stage system and provides additional safety during cryogenic operation: If the 300\,l/s TMP fails, the smaller TMP can maintain a vacuum level good enough to avoid too much frozen gas on the cryogenic surfaces; a failure of the smaller pump does not cause any immediate problem either. The room-temperature base pressure in the TTU is determined by the limited vacuum conductance across the whole system, where some components cannot be baked due to the use of the cryogenic vacuum grease Apiezon N which starts to flow at 42$\,^\circ$C. Nonetheless, after turning on the cryocooler, the large inner surfaces cryosorb residual gas, quickly reducing the pressure to UHV levels. As for the ion-trap chamber, the first and second stage are differentially separated from the vacuum chamber and the vacuum conductance to the segment~III tube is rather small. Therefore, to enhance pumping there, an ion-getter pump is installed. It can be separated from the main vacuum during activation by means of a DN63CF gate valve. The main residual gas in the space between the chamber wall and the first cryogenic stage is H$_2$, which is removed by the getter pump with a speed of 200\,l/s.
Cryopumping at the ion trap is further enhanced by gluing approximately one gram of activated charcoal to a container placed inside the second stage. It acts as a strong getter at cryogenic temperatures due to the surface area of its pores, which can reach thousands of square meters per gram \cite{day_basics_2007}.

It is also possible to close a DN100CF gate valve between the 300\,l/s TMP and the cryogenic system, which allows switching off all mechanical pumps during a measurement, eliminating their vibrations. However, measurements have shown that pump vibrations are also well suppressed by our cryogenic system (see Sec. \ref{sec:VM}).

\subsection{Cryocooler}
We use a model RP-082 pulse-tube cryocooler from Sumitomo Heavy Industries \footnote{Model and supplier are just given for reference. This should not be considered as a recommendation.}, with an F-70H compressor driving it through 20\,m long supply and return lines. The manufacturer specifies vibration levels of 7\,\micro m and 9\,\micro m for its first and second stages, respectively. A thermal load of 40\,W can be refrigerated by the first stage at 45\,K, and 1\,W by the second stage at 4.2\,K. The rotary valve and drive are directly attached to the cold head without vibration decoupling. It only operates in upright orientation, therefore segment~I, where it is mounted, is vertically oriented.

Higher thermal loads raise the achievable temperatures. Our target temperatures were around 45\,K and 4.5\,K for the first and second stage respectively, to reach long HCI storage times on the order of one hour. A low temperature of the first stage allows for a lower temperature of the second one. The presented system does not necessarily rely on the employed cryocooler, and small mechanical modifications of segment~I would accommodate other models with less vibrations or larger refrigeration powers.

\subsection{Thermal transfer unit}\label{sec:TL}
The TTU bridges a horizontal distance of 1.4\,m between the cryocooler and the ion trap. It is vacuum-insulated from the room-temperature chamber. Long, heavy and stiff nested sections carry the heat in each segment. Vibration-decoupling soft links suppress transmission of vibrations between them. The first (outer) stage serves as a heat shield for the second one for both radiation and conductive heat loads. Owing to the distance they have to bridge, the cold stages have large surface areas and consequently a high radiation load. To minimize this load, most of the copper parts are plated with a 10\,\micro m silver layer as a diffusion barrier and a 0.5\,\micro m layer of gold, which prevents surface oxidation and provides a low radiation-emission coefficient.

In segments II and III, low thermal conductivity stainless-steel spokes of 2\,mm diameter are used to suspend the TTU from the room-temperature vacuum chambers as well as the second stage of the TTU from the first one. The spokes are made as long as possible within the given geometry to reduce thermal conduction. In contrast, thermal conduction along the TTU is enhanced by making it as short (and large in cross section) as possible and by using OFHC copper with purities between 4N and 5N (99.99 and 99.999\,$\%$, respectively). Furthermore, most of the copper parts were vacuum-annealed to partially recrystallize them, further increasing their residual-resistivity ratio (RRR) and thermal conductivity $\lambda$\cite{mimura_precise_1997}. Thereby, $\lambda$ values of thousands of Wm$^{-1}$K$^{-1}$ can be achieved\cite{woodcraft_recommended_2005}, corresponding to RRR values of a few hundreds. For parts that were brazed together, such as the first stage of segment I and III, the annealing procedure was carried out simultaneously with the vacuum brazing at a temperature beyond 850\,$^\circ$C. Annealing was performed individually for elements which were bolted together, such as the bars for the pendulum. These were vacuum-annealed at a temperature of only 450\,$^\circ$C over a period of 2 to 3\,hours. Annealing under oxygen atmosphere would generally lead to even higher RRR values due to oxidation of magnetic impurities\cite{woodcraft_recommended_2005}, but safety concerns precluded this. To reduce the thermal boundary resistance between different surfaces which are clamped together, we use thin films of the cryogenic vacuum grease Apiezon N to close microscopic irregularities\cite{salerno_thermal_1994,gmelin_thermal_1999}.

\subsubsection{Segment I}
A 220\,mm-long first-stage hollow octagon is vacuum brazed from eight prismatic copper bars of 10\,mm thickness (see Fig.\,\ref{fig:TL}). It is attached to the first stage of the cryocooler and houses the second stage. It ends inside the six-way cross, where flexible copper links connect it to the next segment of the TTU. The second stage of the pulse-tube cryocooler is directly connected by means of flexible copper links to the next segment. 

\subsubsection{Segment II}\label{sec:TL_SII}
\begin{figure}
	\includegraphics[width=0.95\linewidth]{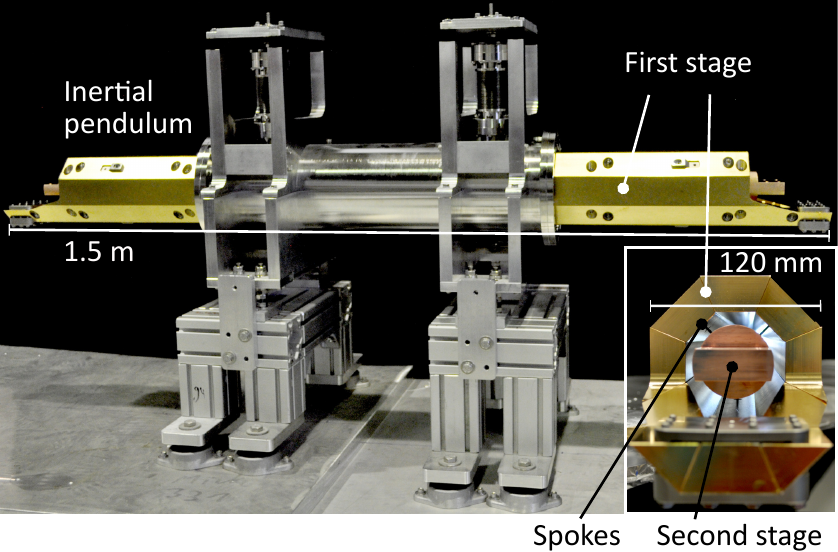}
	\caption{Photograph of segment II prior to the installation through the noise insulating wall\label{fig:pendulum}. The pendulum hangs freely inside the vacuum chamber, suspended by two spokes from the top flanges of two vertical DN40CF bellows. The inset shows a view along the pendulum exposing the spokes holding the second stage from the first one.}
\end{figure}

The first stage of the pendulum (see Figs.\,\ref{fig:TL} and \ref{fig:pendulum}) is also a hollow octagon made by bolting together eight $\sim1.4$\,m-long and 25\,mm-thick copper bars of trapezoidal cross section. On top of it, two stainless-steel inserts hold two vertical 150\,mm long stainless-steel spokes that are screwed by their other end to the vacuum side of DN40CF blind flanges. These close the vertical DN40CF bellows connected to the DN40CF ports of the horizontal vacuum chamber and are externally supported by the structure described in Sec. \ref{sec:PS}. Inside the octagon, the second-stage rod is held by six counteracting spokes of 110\,mm length. The rod, made from OFHC 4N5 (99.995\,$\%$) copper, has a diameter of 50\,mm and a length of 1.34\,m.

\subsubsection{Segment III}\label{sec:TL_SIII}
\begin{figure}
	\includegraphics[width=0.95\linewidth]{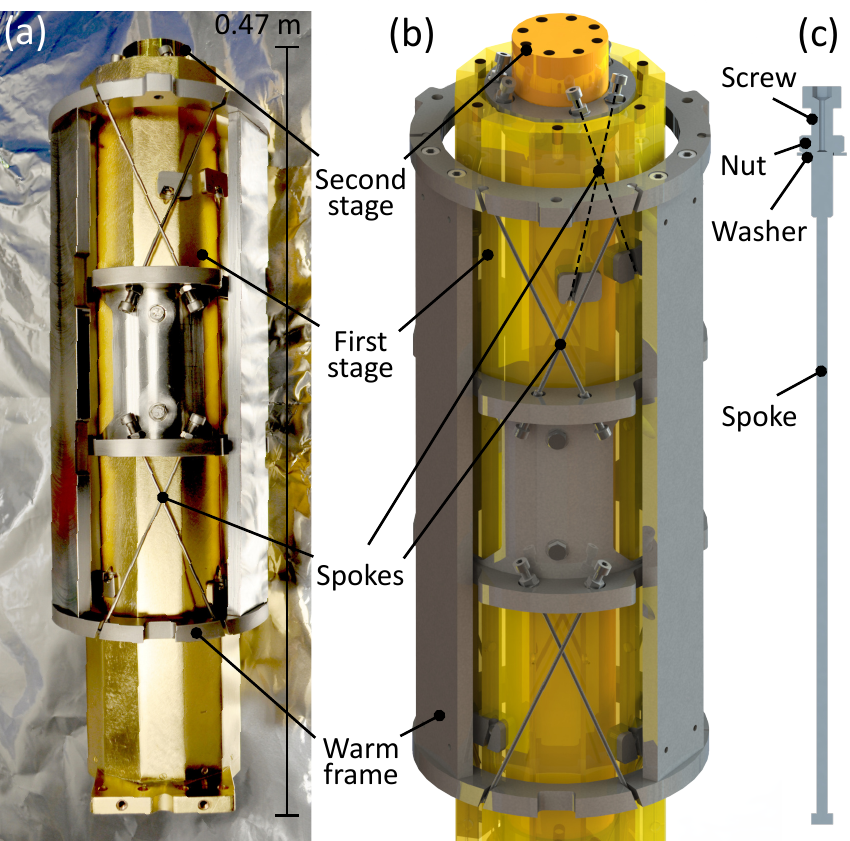}
	\caption{(a) Photograph of segment III of the TTU\label{fig:VP}. The first stage is suspended by spokes from a room-temperature frame. In the 3D model (b) the first stage is displayed as transparent to show the inside mounting of the second stage, also by means of spokes. (c) shows a cross section of a vacuum-brazed spoke. See text for further details.}
\end{figure}

The two stages of segment III are supported by a room-temperature frame made of stainless steel (see Figs.\,\ref{fig:TL} and \ref{fig:VP}) resting inside of its DN160CF vacuum tube (see Sec. \ref{sec:PS}). From the frame, twelve 120\,mm-long spokes are arranged in counteracting pairs to hold the hollow 0.45\,m-long first-stage dodecagon, vacuum brazed from twelve trapezoidal copper bars. The spokes are vacuum-brazed into fitting bores drilled along the axis of Allen-head screws. A nut on the thread of each screw is used to adjust position and tension (see Fig. \ref{fig:VP}). The low-temperature ends of the spokes are held by a stainless-steel girdle screwed to the first stage. The warm ends are hooked into the room-temperature frame. From the first stage a 50\,mm-diameter second-stage copper rod with a length of 0.44\,m, made from OFHC 4N5(99.995$\%$) copper, is suspended again by means of twelve spokes with lengths between 80\,mm and 90\,mm in a similar arrangement.

\subsubsection{Flexible copper links}
\begin{figure}
	\includegraphics[width=0.95\linewidth]{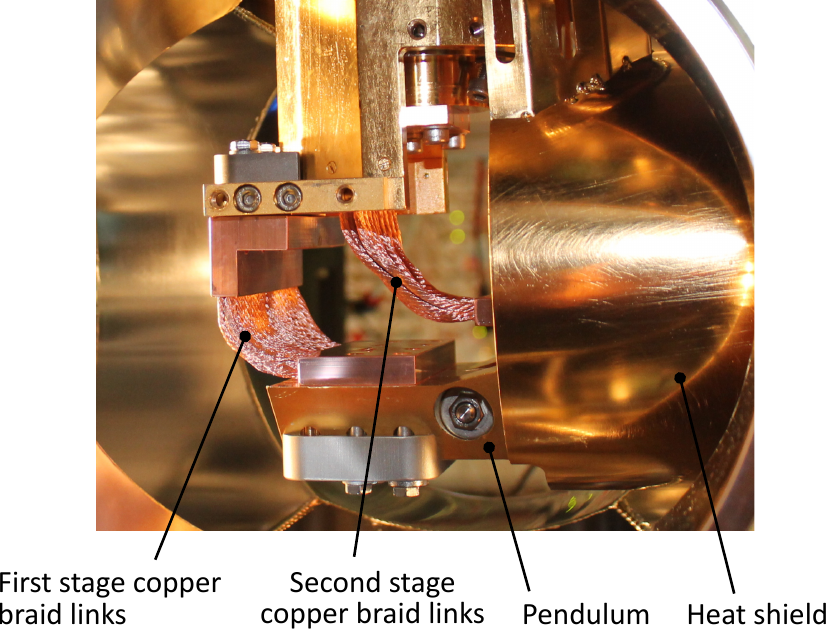}
	\caption{Flexible copper braid connections between segments II and III. The gold-plated heat shield (see Fig.\,\ref{fig:HS}) that protects the braids from room-temperature radiation is temporarily removed to allow access.}\label{fig:CL}
\end{figure}

Vibration transmission from the pulse tube along the TTU is suppressed by flexible copper links mounted between the three segments. We use customized links from Technology Applications Inc., based on OFHC copper-wire braids (UltraFlex)\footnote{Model and supplier are just given for reference. This should not be considered as a recommendation.} with an estimated purity between 4N5 and 4N8. The braids are cold-pressed into copper plates of at least 4N purity. This did not seem to reduce their conductance at operating temperature\cite{dhuley_thermal_2017}. The braids have a diameter of 2.54\,mm and lengths between 35\,mm and 68\,mm. The links for the second stages use 32 individual ropes each, whereas for the first stage connections 68 and 64 ropes on the pulse-tube and trap side are used, respectively. We vacuum annealed them at 550\,$^\circ$C for 6\,hours to increase the RRR value and soften them. Higher annealing temperatures should in principle lead to even better results, but tests on a sample indicated partial fusing of the individual copper strands in the ropes, making them stiffer. Up to 8 bolts per end plate ensured a large number of contact points, used with molybdenum washers for compensation of the differential thermal contraction of copper and stainless-steel parts. As elsewhere, a thin Apiezon N layer was used between contact surfaces. Some of the fittings have holes for mounting temperature sensors. Threaded inserts (Helicoil Nitronic 60) were used in the copper end plates having a high resistance to galling. This permits application of strong force for good thermal contact. The braids are bent by 90 degrees, with several millimeters of excess length and the outer ropes are longer than the inner ones to prevent tension. The links installed beneath segment III are shown in Fig.\,\ref{fig:CL}. 

\subsection{Heat shields}
\begin{figure}
	\includegraphics[width=0.8\linewidth]{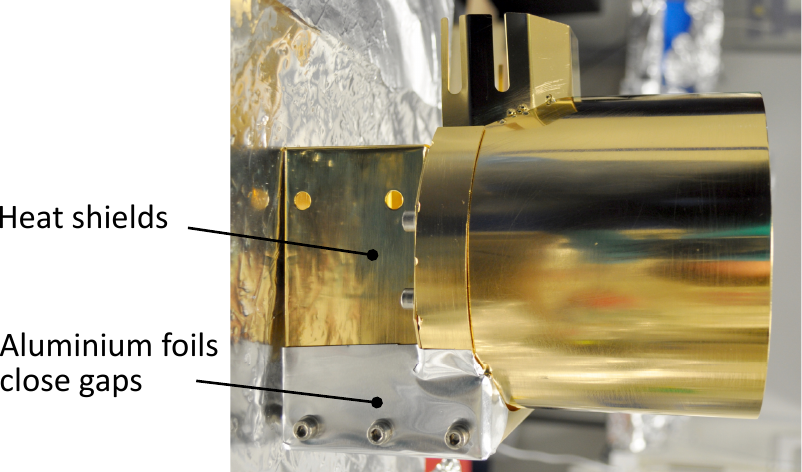}
	\caption{Photograph of a copper-sheet heat-shield assembly enclosing the flexible copper braids in the second cross\label{fig:HS}. Thick aluminum foil is used to cover gaps in order to reduce thermal radiation from the vacuum chamber reaching the second stage. See text for further details.}
\end{figure}

In each cross, the flexible links and sensors are enclosed within assemblies of bent 1\,mm-thick copper sheets (3N5 99.95\,$\%$) as shown in Fig.\,\ref{fig:HS}. Thermally connected to the first stages of segment I and III, respectively, they shield parts within from room-temperature black-body radiation. Furthermore, the sensors require shielding to exclude perturbations due to radiation directly heating them. To prevent transmission of vibrations through these heat shields, care was taken to avoid contact with the pendulum of segment II.

\subsection{Temperature sensors}
Six sensors monitor temperatures at various locations (see Fig.\,\ref{fig:overview}). Four of them are permanently installed before and after the first and second stage flexible links in the first cross. Here we employ calibrated Cernox-type sensors, since these are magnetic-field insensitive and operate in closer proximity to the magnetic field of the EBIT used for HCI production. These sensors measure the temperature gradient across the flexible links. Temporarily, we also installed two silicon-diode temperature sensors at the beginning of the first and second stage of segment III to measure the temperature gradient up to the ion trap where two more silicon sensors are permanently installed on the first and second stage inside the respective temperature stage enclosures. Since these are completely enclosed there, thermal radiation from a warmer stage is blocked. Thin cryogenic 0.127\,mm-diameter (Cernox sensors) and 0.202\,mm-diameter (silicon diodes) phosphor-bronze wires are used for connecting the sensors to the vacuum-feedthrough flange in a 4-wire arrangement. The wiring was thermally anchored with vacuum-compatible Kapton tape to the temperature stages and had sufficient length in between for heat-load reduction. 

\subsection{Restricting the motion of the optical table}
	\begin{figure}
	\includegraphics[width=0.95\linewidth]{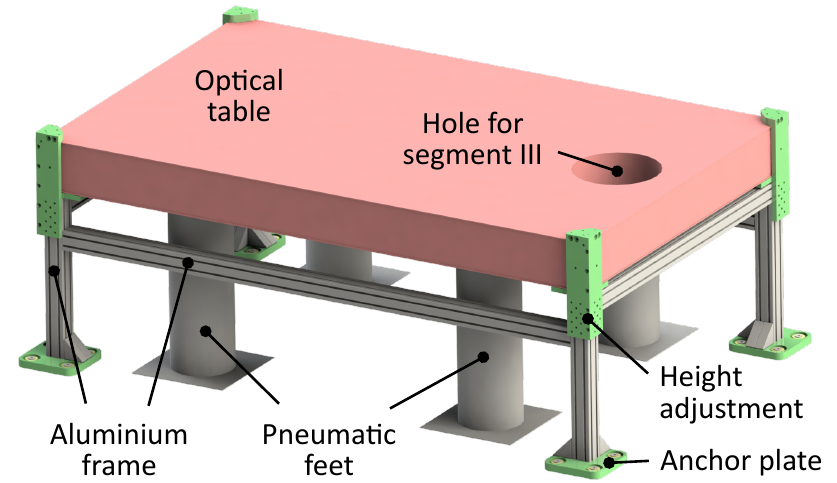}
	\caption{Optical table and frame restricting its motion range \label{fig:frame} within adjustable mechanical stops. The four corner pillars are firmly anchored to the concrete floor; horizontal bars stiffen the frame.}
\end{figure}
For suppression of vibrations, free relative motion of the different segments, bound together by bellows, is needed. Owing to the small gaps between both, the cryogenic stages themselves in the different segments and the vacuum chamber walls, undesired contact can occur when the floating optical table moves by more than a few millimeters. To avoid thermal shorts we restrict the motion of the optical table to less than $\pm 1$\,mm in each direction by means of a strong aluminum frame, shown in Fig.\,\ref{fig:frame}. It is rigidly anchored to the concrete floor and has adjustable mechanical stops that can, if needed, withstand the force exerted by the pneumatic legs, which are equipped with position regulators that keep the table within those limits.

\subsection{Installation}
The TTU was assembled within a few days after completion of its different components. First, segment II (see Fig.\,\ref{fig:pendulum}) was mounted. Second, the fully-assembled segment I and the six-way cross of segment III (second cross) were attached with the inner edge-welded bellows to segment II. Temporarily, we used for the second cross a similar support structure as for the first cross, shown in Fig.\,\ref{fig:overview}. Then, the vertical element of section III, (i.\,e. the DN160CF tube with its inner elements) was passed through the hole in the optical table (see Fig.\,\ref{fig:frame}) and connected to the top of the second cross. Following that, the ion trap vacuum chamber was installed on top of it and rigidly attached to the optical table. The outer edge-welded bellows were mounted to the crosses and anchored to the pillars fixed on the floor. The TTU was completed by installing the flexible links inside the crosses through the DN200CF flanges. At the same time the temperature sensors and the heat shielding enclosures were installed. Finally, the optical table was floated, lifting the second cross from its temporary support structure.

\section{Performance characterization}

\subsection{Vacuum}
\begin{figure}
	\includegraphics[width=0.9\linewidth]{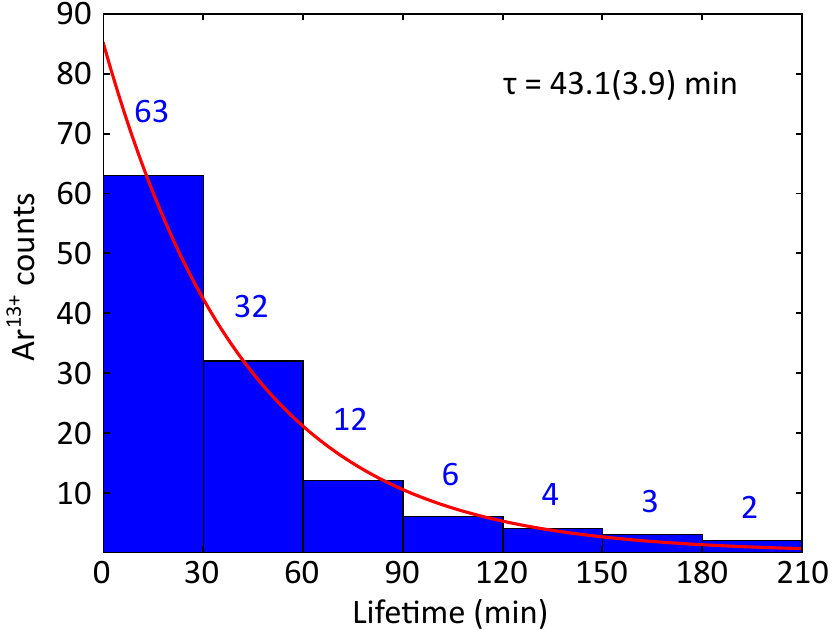}
	\caption{Lifetime measurement of highly charged Ar$^{13+}$\label{fig:LT}. The data visualized in the histogram yield a lifetime of $43.1(3.9)$\,min, evaluated by a maximum likelihood estimate. The corresponding exponential decay function is shown in red.}
\end{figure}

The residual gas pressure between the vacuum chamber and the first stage is measured by a hot-cathode ion gauge installed at the first cross behind a DN40CF elbow to prevent thermal radiation from the gauge reaching the first stage. Without cryopumping, the pressure reaches the 10$^{-8}$\,mbar level after a few days of pumping. With the cryocooler switched on, the pressure drops to $3 \times 10^{-10}$\,mbar at the end of the cooldown. Inside the second stage no pressure gauges are installed, but the vacuum pressure can be estimated by the lifetime of the ions stored in the Paul trap. At temperatures below 5\,K molecular hydrogen is the dominant residual gas. As in other experiments, singly-charged Be$^+$ ions under excitation by the 313\,nm Doppler-cooling laser can undergo a chemical reaction forming BeH$^+$ \cite{roth_ion-neutral_2006,sawyer_reversing_2015}. With a large cloud of stored Be$^+$, the residual gas pressure can thus be estimated. At the National Institute of Standards and Technology (NIST), a pressure of $5\times10^{-11}$\,mbar was estimated from a Be$^+$ lifetime of 2.2\,hours when the ions were continuously excited with a saturation parameter of 0.3\cite{sawyer_reversing_2015}. Before installing the activated charcoal getter in our system, we also observed BeH$^+$ formation on such time scales a few days after a cooldown, indicating a degradation of the initially much better pressure. This is in agreement with our experiences from CryPTEx\cite{schwarz_cryogenic_2012,schmoger_coulomb_2015}. The increasing amount of adsorbed residual gas on the cryogenic surfaces reduces their pumping speed. Short warmup cycles had to be performed every evening to restore initial vacuum conditions.

For our new system a short warmup cycle takes approximately one to three hours, allowing the second stage to typically reach 20 to 30\,K. Shortly after switching off the cryocooler, the pressure measured by the hot-cathode ion gauge typically increases to the $10^{-7}$\,mbar range, then slowly returning to below $10^{-8}$\,mbar over a period of typically one hour as the TMPs remove the desorbed gas. After passing through a minimum, the pressure would raise again if the warmup were continued. Here, we switch the cryocooler back on. 

After installation of the getter, we have never seen BeH$^+$ formation and trap single Be$^+$ routinely for a whole week without losing them. This is due to the much greater adsorption capacity of the charcoal compared to the mostly polished metal surfaces. Be$^+$ ions are only lost if the cooling laser is turned off for many days. Assuming a Be$^+$ ion exposed to the same 313\,nm cooling laser saturation as for the aforementioned NIST experiment, for about 10\,hours per day and for a conservative estimate of five days storage time without reaction, we can place an upper limit of $2\times10^{-12}$\,mbar on our pressure. A more sensitive and efficient vacuum probe is an HCI on which charge exchange (CX) reactions with hydrogen can be observed (see, e.\,g., \cite{schmoger_kalte_2017} and references therein). Following this approach, we stored a single highly charged Ar$^{13+}$ ion, sympathetically cooled by a single laser-cooled Be$^+$ ion, in our Paul trap and measured the lifetime of this two-ion crystal. CX is immediately observable through the temporary decrystallization of the Coulomb crystal or through loss of one or both ions from the trap. Fig.\,\ref{fig:LT} shows the recorded lifetimes of a set of 122~two-ion crystals within two weeks after a warmup cycle. By a maximum likelihood estimate we obtain a mean lifetime of $\tau = 43.1(3.9)$\, min.  In the classical over-barrier model (see, e.\,g., \cite{niehaus_classical_1986}), the critical distance for CX between H$_2$ and the HCI is proportional to $\sqrt{q}$, where $q$ is the HCI charge. According to the Langevin model\cite{langevin_formule_1905}, the residual-gas particle density is given by
\begin{equation}
	\label{eq:LM}
	n = \frac{1}{\tau \cdot k_L}
\end{equation}
where $\tau$ is the HCI lifetime and $k_L = q/(2\varepsilon_0) \sqrt{\alpha/\mu}$ is the Langevin rate coefficient, $\varepsilon_0$ the electric constant, $\alpha = 4 \pi \varepsilon_0 (0.787\times10^{-24})$\,cm$^3$ the H$_2$ polarizability, and $\mu$ the reduced mass of the HCI-H$_2$ system. For (Ar$^{13+}$, H$_2$) $k_L$ has a value of $1.957\times10^{-8}$\,cm$^3$s$^{-1}$ (see reference \cite{schmoger_kalte_2017}). The pressure is then given by
\begin{equation}
	\label{eq:p}
	p = n \cdot k_B \cdot T
\end{equation}
with the Boltzmann constant $k_B$ and the residual gas temperature $T$, which is assumed to be in thermal equilibrium with the second stage. Based on Eqs.\,\ref{eq:LM} and \ref{eq:p}, we can give an upper limit for the cryogenic vacuum pressure at the Paul trap inside the second stage, since the measured lifetime is only a lower limit for the actual lifetime of the described loss process. Additional non-negligible loss processes, e.\,g., due to collisions or imperfections of the trap drive electronics, which cause a configuration change of the crystal or a temporary heating, as well as collisions with ballistic room-temperature gas entering the trap region through the open laser access ports \cite{leopold_cryogenic_nodate}, shorten the lifetime to the above measured value. Hence, we derive an upper XHV pressure limit of $p<1.26\,(-0.11/+0.12)\, \times 10^{-14}$\,mbar at $T=4.6$\,K, strongly indicating operation on the $10^{-15}$\,mbar level as expected for such a cryogenic setup. Arguably, the cryogenic vacuum pressure could be improved and the HCI lifetimes prolonged by closing some of the laser ports in the first stage with windows, though this was not done here due to concerns about birefringence and potentially longer initial pumpdown times due to the decreased pumping conductance. With all laser ports open, $0.01\,\%$ of the total solid angle exposes the ion to room temperature \cite{leopold_cryogenic_nodate}.

As mentioned above, the HCI lifetime and the vacuum pressure degrade over time. Thus, the measured lifetime is an average over the period of two weeks after a warmup cycle. Significantly longer HCI lifetimes are observed shortly after such a cycle. It is a pragmatic choice when to perform a short warmup with the tradeoff of losing parts of a day due to the cycle time.

\subsection{Temperatures}

\subsubsection{Cooldown}
\begin{figure}
	\includegraphics[width=1.\linewidth]{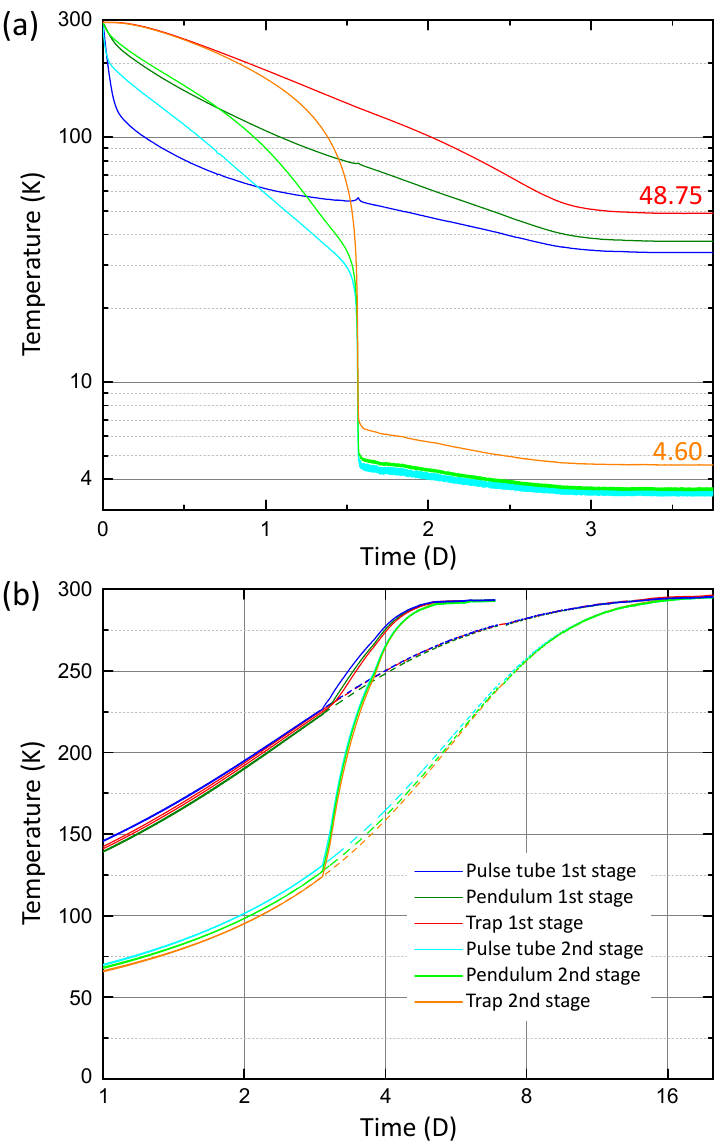}
	\caption{Cooldown (a) and warmup (b) behavior\label{fig:cooldown}. (a) Temperature evolution measured by six sensors. At the ion trap, the steady-state temperatures are 48.8\,K and 4.6\,K on the first and second stage, respectively. The pendulum temperature is measured at the first cross behind the flexible copper links. (b) A warmup of the system after switching off the pulse tube takes more than 14 days (dashed lines). A faster warmup is aided by letting a small amount of helium into the vacuum chamber after about 3 days for increased thermal coupling between the temperature stages (solid lines). Then, the system reaches room temperature in less than two days. Note the different scales in (a) and (b). See text for further details.}
\end{figure}

It is important to consider that before a cooldown the pressure inside the cold stages of such a nested design is significantly worse than the pressure measured by an outside gauge which is closer to the pumps. Therefore, a cooldown is usually not started before reaching a vacuum in the $10^{-8}$\,mbar range as measured by the gauge in the first cross. This reduces the amount of freezing residual gas and prolongs the time to the first warmup cycles in order to restore initial vacuum conditions. Furthermore an increase of the emissivity of the cryogenic surfaces due to adsorbed layers of residual gas is prevented.

We continuously log the temperatures of the different segments of the TTU with sensors mounted at the locations shown in Fig.\,\ref{fig:overview}. We reach 4.6 and 49\,K with cooldown times of 72 and 36\,hours for the first and second stage, respectively (see Fig.\,\ref{fig:cooldown}\,(a)). After cooling down for about 36\,hours the second stage temperatures quickly decrease to values near steady-state. This effect is due to the freezing of degrees-of-freedom in the copper, that result in a much-reduced specific heat capacity and an increased thermal conductivity below 40\,K, especially for the annealed parts. In many cases experiments can already proceed 36\,hours after the start of the cooldown. However, thermal contraction could still change the ion trap alignment with the laser. In practice, cooling times of three days over the course of a weekend were a sensible option. We usually maintain cryogenic temperatures over months. In contrast, to restore UHV/XHV conditions after venting in a room-temperature setup requires several weeks of cumbersome high-temperature bake-out procedure to reach pressures in the $10^{-11}$\,mbar range.

\subsubsection{Warmup}
Before venting the system to perform work in the vacuum chamber, it has to warm up above the dew point (around 8\,$^\circ$C at a relative humidity of 40\,$\%$ at 22\,$^\circ$C). Due to the good thermal insulation of the cold stages, the whole system is basically only warmed up through the switched-off pulse tube, and this process takes two weeks (see Fig.\,\ref{fig:cooldown}\,(b)). To speed it up, a small amount of helium gas can be leaked into the chambers, increasing the pressure to $10^{-2}$\,mbar or even more, but yet keeping the vacuum chamber temperature around 12\,$^\circ$C -- just above the dew point to avoid water condensation on the laser windows. Since He does not freeze out and is a light atom, it ensures a fast heat transfer, causing a cooling of the vacuum chamber wall while the inside temperatures quickly rise. To keep the helium from being pumped away, the gate valve to the TMPs is closed. To exclude a damaging overpressure due to the release of cryosorbed gas, we first warm up the two stages to approximately 100 and 200\,K under pumping before closing the gate valve and letting helium into the chamber.

\subsubsection{Thermal budget}\label{sec:budget}
We performed thermal calculations, using established methods as described, e.\,g., in \cite{ekin_experimental_2006}, taking into account thermal resistances, conductances, and radiation for optimizing the heat load and cryogenic temperatures. These calculations are fairly consistent with the performance of the built systems.

The thermal resistances $R_{th}$ of the various rigid copper elements were minimized along the TTU. They were derived from Fourier's law of heat conduction at a constant nominal temperature (of 4 and 40\,K for the second and first stage, respectively), yielding
\begin{equation}
	\label{eq:R_th}
	R_{th} = \frac{l}{A \cdot \lambda \left( T \right) }\, ,
\end{equation}
depending on the cross section $A$, the length $l$ and the material-specific and temperature-dependent thermal conductivity $\lambda$. We maximized the cross sections of the various parts given the geometric constraints of the system; the lengths were determined by the size of the apparatus. Extremely pure copper was used for high $\lambda$ values with vacuum annealing applied wherever practicable to further increase the conductivity at low temperatures.

The conductances of the flexible copper links can be roughly estimated with the capacity map of the pulse tube\cite{noauthor_rp-082b2_2018} yielding the total thermal loads of the two temperature stages as a function of their steady-state temperatures. The temperature gradients across the links are measured in the first cross by temperature sensors on their end plates (see Fig.\,\ref{fig:overview}). For the first stage we observe a thermal load of 22\,W (with the ion trap chamber \cite{leopold_cryogenic_nodate} installed) and a temperature gradient of 3.8\,K, leading to a conductance of $5.8\,\textrm{W}\,\textrm{K}^{-1}$. For the second stage we obtain a value of $0.8\,\textrm{W} / 0.15\,\textrm{K} = 5.3\,\textrm{W}\,\textrm{K}^{-1}$. Since the links attached to the first and second stage have comparable lengths, but differ in the number of copper ropes by a factor of two, these measurements imply that the conductivity of the copper material at 3.5\,K is twice as high as its value at 33.8\,K. This indicates RRR values between 500 and 1200\cite{simon_properties_1992} -- a substantial improvement upon a measured value of 77 for an un-annealed strap\cite{dhuley_thermal_2017} resulting in a five- to tenfold increase of the thermal conductivity at 4\,K. The conductances of the flexible links in the second cross can be estimated by scaling the values of the first links regarding the slightly different lengths. The estimated values $R_{th}$ for the different parts of the TTU are listed in Tab.\,\ref{tab:ThermResistances}.

\begin{table}
	\caption{Estimated thermal resistances $R_{th}$ of the connecting copper elements of the TTU. An RRR of about $1000$ is assumed for the annealed copper of the rigid parts with purities between 4N5 and 5N; thermal conductivities $\lambda\left( T \right)$ are taken from \cite{johnson_compendium_1961,simon_properties_1992}. For the flexible links $R_{th}$ was estimated as described in the text. The nominal temperatures assumed for the two temperature stages are 4 and 40\,K, respectively.\label{tab:ThermResistances}}
	\begin{tabular}{lcccc}
		\hline
		\hline
		Stage & Element & $\lambda$  				& $R_{th}$ \\
		&	& (W\,cm$^{-1}$\,K$^{-1}$) & (K\,W$^{-1}$)	\\
		\hline
		2$^{\mathrm{nd}}$ & 1$^{\mathrm{st}}$ flexible link & - & 0.19 \\
		2$^{\mathrm{nd}}$ & Pendulum rod & 70 & 0.10  \\
		2$^{\mathrm{nd}}$ & 2$^{\mathrm{nd}}$ flexible link & - & 0.25 \\
		2$^{\mathrm{nd}}$ & Vertical rod & 70 & 0.03 \\
		1$^{\mathrm{st}}$ & 1$^{\mathrm{st}}$ vertical shield & 20 & 0.03 \\
		1$^{\mathrm{st}}$ & 1$^{\mathrm{st}}$ flexible link & - & 0.17 \\
		1$^{\mathrm{st}}$ & Pendulum shield & 20 & 0.09 \\
		1$^{\mathrm{st}}$ & 2$^{\mathrm{nd}}$ flexible link & - & 0.11 \\
		1$^{\mathrm{st}}$ & 2$^{\mathrm{nd}}$ vertical shield & 20 & 0.01 \\
		\hline
		\hline
	\end{tabular}
\end{table}

Estimates of the total heat load onto the two cryocooler stages include heat conduction by the spokes, connecting parts of different temperatures, and thermal radiation from warmer surfaces. The heat flow through a spoke from a thermal reservoir at temperature $T_2$ to another reservoir at $T_1$ is given by
\begin{eqnarray}
	\label{eq:Q_con}
	\dot Q_{cond} & = & \frac{A}{l} \int\limits_{T_1}^{T_2} \lambda\left(T\right)  \,\mathrm{d}T \\
	& = & \frac{A}{l} \left( \, \int\limits_{4K}^{T_2} \lambda\left(T\right)  \,\mathrm{d}T - \int\limits_{4K}^{T_1} \lambda\left(T\right)  \,\mathrm{d}T \right) .
\end{eqnarray}
The thermal conductivity integral values for stainless steel, 3060\,W/m at T$_2 = 300$\,K and 82\,W/m at T$_1 =40$\,K, were taken from \cite{ekin_experimental_2006}. We minimized the heat flow through the spokes by making them as long as practically possible, using lengths between 79\,mm and 150\,mm, yielding 1.09\,W and 60\,mW onto the first and second stage, respectively (see Tab.\,\ref{tab:heatinput}).

\begin{table}
	\caption{Heat load by thermal conduction through the spokes. The elements given in the first column, being at the nominal temperature $T_1$, receive the heat input, given in the last column, by a certain number of spokes of a certain length from the warmer reservoir at $T_2$. The vertical rod has two different types of spokes, (I) and (II), which differ in their length by about 10\,mm. \label{tab:heatinput}}
	\begin{tabular}{lccccc}
		\hline
		\hline
		Element& $T_2$&$T_1$& Number & Length & Heat input\\
		 &  (K)&  (K)&   & (mm) &  (W)\\
		\hline
		Pendulum shield & 300 & 40 & 2 & 150 & 0.13 \\
		Vertical shield & 300 & 40 & 12 & 120 & 0.96 \\
		Total 1$^{\mathrm{st}}$\,stage & - & - & 14 & - & 1.09 \\
		\hline
		Pendulum rod & 40 & 4 & 6 & 100 & 0.02 \\
		Vertical rod (I)& 40 & 4 & 6 & 79 & 0.02 \\
		Vertical rod (II)& 40 & 4 & 6 & 89 & 0.02 \\
		Total 2$^{\mathrm{nd}}$\,stage & - & - & 18 & - & 0.06 \\
		\hline
		\hline
	\end{tabular}
\end{table}

\begin{table}
	\caption{Estimated steady-state heat loads of the cryostat without the ion trap.\label{tab:load}}
	\begin{tabular}{lcc}
		\hline
		\hline
		& 1$^{\mathrm{st}}$\,stage & 2$^{\mathrm{nd}}$\,stage\\
		&(W)&(W)\\
		\hline
		Black-body radiation & 8.8 & 0.001\\
		Conduction through spokes & 1.09 & 0.06 \\
		Conduction through sensor wiring & 0.01 & $<$0.01\\
		Total & 9.9 & 0.06\\
		\hline
		\hline
	\end{tabular}
\end{table}

The thermal radiation heat loads onto the first and second stages were estimated by approximating the cryogenic system as three long, concentric cylinders. For diffuse reflection, the Stefan-Boltzmann law and the cylindrical geometry yield
\begin{equation}
	\label{eq:Q_rad}
	\dot Q_{rad} = \frac{A_1 \, \sigma \, \left( T_2^4 - T_1^4 \right) }{\frac{1}{\epsilon_1} + \frac{A_1}{A_2} \left( \frac{1}{\epsilon_2} - 1 \right) }\, ,
\end{equation}
where the power $\dot Q_{rad}$ is emitted from the outer surface with area $A_2$ at temperature $T_2$ with emissivity $\epsilon_2$, and is received by the inner surface $A_1$ at temperature $T_1$ with emissivity $\epsilon_1$. $\sigma$ is the Stefan-Boltzmann constant. For specular reflection $A_1 = A_2$ holds in the denominator\cite{ekin_experimental_2006}, resulting in only minor corrections. For the nominal temperatures of 300\,K (vacuum chamber), 40\,K (first stage), and 4\,K (second stage) and estimated emissivities of $\epsilon_1 = \epsilon_2 = 0.05$ for the electropolished stainless-steel vacuum chamber and the mainly gold-plated copper parts, we calculate radiative heat loads of 8.8\,W onto the first stage and 1\,mW onto the second stage.

In Tab.\,\ref{tab:load} the estimated total heat loads of only the cryostat without the ion trap chamber and ion trap are given. The dominant heat load onto the first stage is the thermal radiation, while for the second stage the thermal conduction is dominant. Additional heat loads onto the first and second stages are introduced by thermal radiation from the ion trap chamber, conduction through further spokes and the rf ion trap wiring, and dissipation of rf power \cite{leopold_cryogenic_nodate}. Thus, we reach final temperatures of 33.8\,K at the pulse-tube first stage and of 3.50\,K at the second stage, respectively. These values can be compared with the capacity map of the pulse-tube cryocooler\cite{noauthor_rp-082b2_2018}, roughly yielding heat loads of about 22\,W (first stage) and 0.8\,W (second stage).

\subsection{Stability}\label{sec:VM}

\subsubsection{Decoupling from vibration and noise sources}
\begin{figure}
	\includegraphics[width=0.95\linewidth]{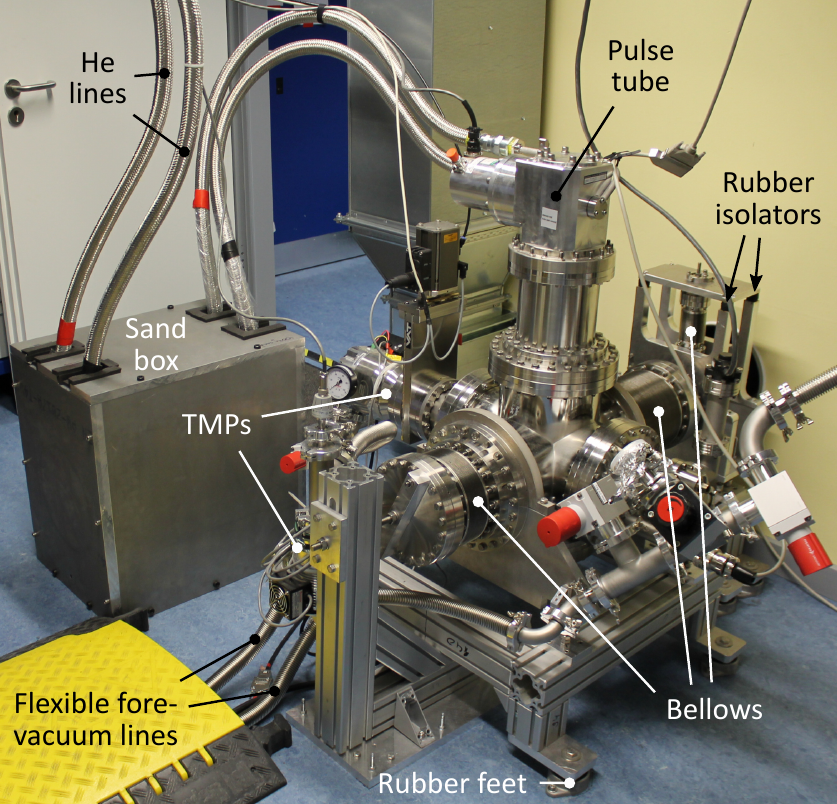}
	\caption{Photograph of the cryostat in the machine room\label{fig:MR}. Vibration sources and decoupling elements are labeled. TMPs -- turbomolecular pumps}
\end{figure}

Sources of vibrations are the pulse-tube cryocooler (1.7\,Hz) and its compressor, the forevacuum scroll pump (around 25\,Hz), the TMPs (1000\,Hz and 1500\,Hz, respectively), and others that couple through the laboratory floor: building vibrations, air-conditioning, water-cooling systems, and fans. All of these also produce acoustic noise coupling through the air and walls. We installed noisy parts in the machine room, which is acoustically insulated from the laser laboratory, achieving a noise reduction of approximately 20\,dB. The machine-room floor has a vibration-damping screed. The compressor is mounted on a vibration-insulation platform on passive pneumatic feet which decouple it from the floor. However, vibrations can still be transmitted along the cryogenic helium lines. To damp such compressor noise we guide the helium hoses through a heavy metal box filled with 100\,kg of fine quartz sand placed right next to the pulse tube (see Fig.\,\ref{fig:MR}). Also, the forevacuum scroll pump rests on a pneumatic vibration-insulation platform, and the flexible stainless steel hose connecting it to the segment~I TMPs is guided through a second metal box with 25\,kg quartz sand. To keep vibrations, arising in segment I by the He gas flow in the pulse tube, from being transmitted towards the ion trap, the three segments are mechanically decoupled. As described in Sec. \ref{sec:CD}, the TTU employs flexible copper links while the vacuum chambers are decoupled by edge-welded bellows with 40 diaphragm pairs each. Segment I rests on rubber feet on the machine-room floor. Segment II with the pendulum acting as a low-pass filter also rests on rubber feet on the floors of both rooms. These feet decouple it from vibrations from the pulse tube and vacuum pumps. The two plates from which the pendulum hangs (as described in Sec. \ref{sec:CD}) are supported by four rubber isolators each (as shown in Figs.\,\ref{fig:overview} and \ref{fig:MR}). Segment III is connected to segment II through the second vibration-decoupling section and attached to the optical table without any rigid connection to the floor. The pneumatically floating optical table with a mass of $> 600$\,kg acts as a second inertial filter for vibration damping. In this way, high-frequency vibrations are efficiently suppressed; low-frequency vibrations are less strongly damped, as shown by the measured vibration spectra (see next section for further details).

\subsubsection{Vibration measurement}
\begin{figure}
	\includegraphics[width=0.95\linewidth]{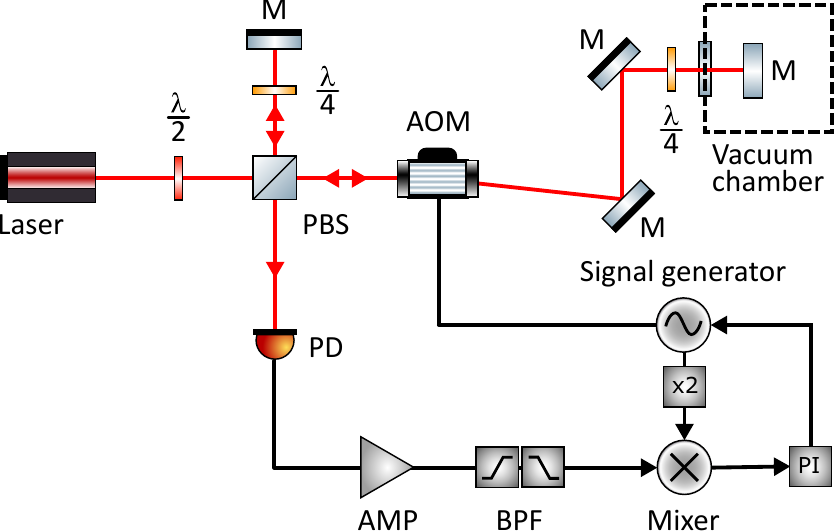}
	\caption{Interferometric setup to measure the vibrations at the ion trap. Two of the three directions were measured with such a closed-loop setup; the third direction was measured in open-loop. See text for further details. AMP -- amplifier, AOM -- acousto-optic modulator, BPF -- band-pass filter, $\lambda / 2$ -- half-wave plate, $\lambda / 4$ -- quater-wave plate, M -- mirror, PBS -- polarizing beamsplitter cube,  PD -- photodiode, PI -- proportional-integral controller.} \label{fig:vibr_setup}
\end{figure}

\begin{figure}
	\includegraphics[width=1.\linewidth]{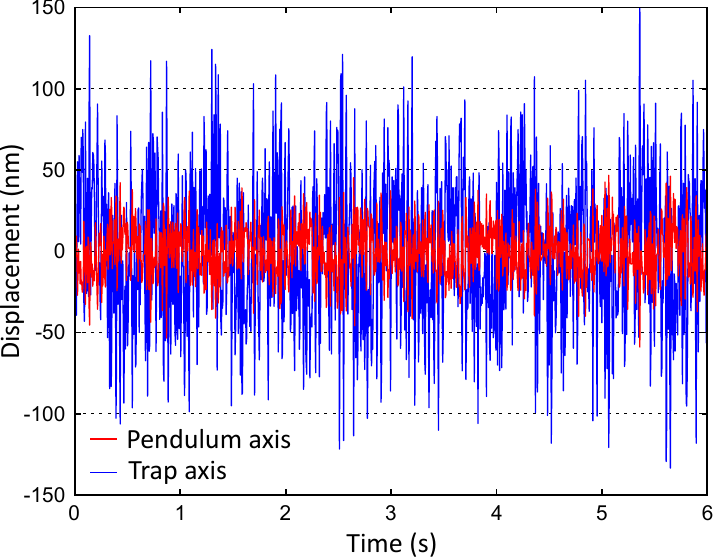}
	\caption{Vibrations in the horizontal plane along the pendulum axis (red) and the perpendicular trap axis (blue)\label{fig:timetrace}. The time traces were recorded with an oscilloscope at a sampling frequency of 200\,Hz. Here, a slow drift was removed by processing the time traces with a cut-off frequency of 1\,Hz. The vibrations along the trap axis had to be reconstructed with the pendulum axis measurement and a vibration measurement taken at $60^\circ$ with respect to the trap axis. A division by $\cos(60^\circ)$ results in a deteriorate accuracy. See text for further details.}
\end{figure}

\begin{figure*}
	\includegraphics[width=0.95\linewidth]{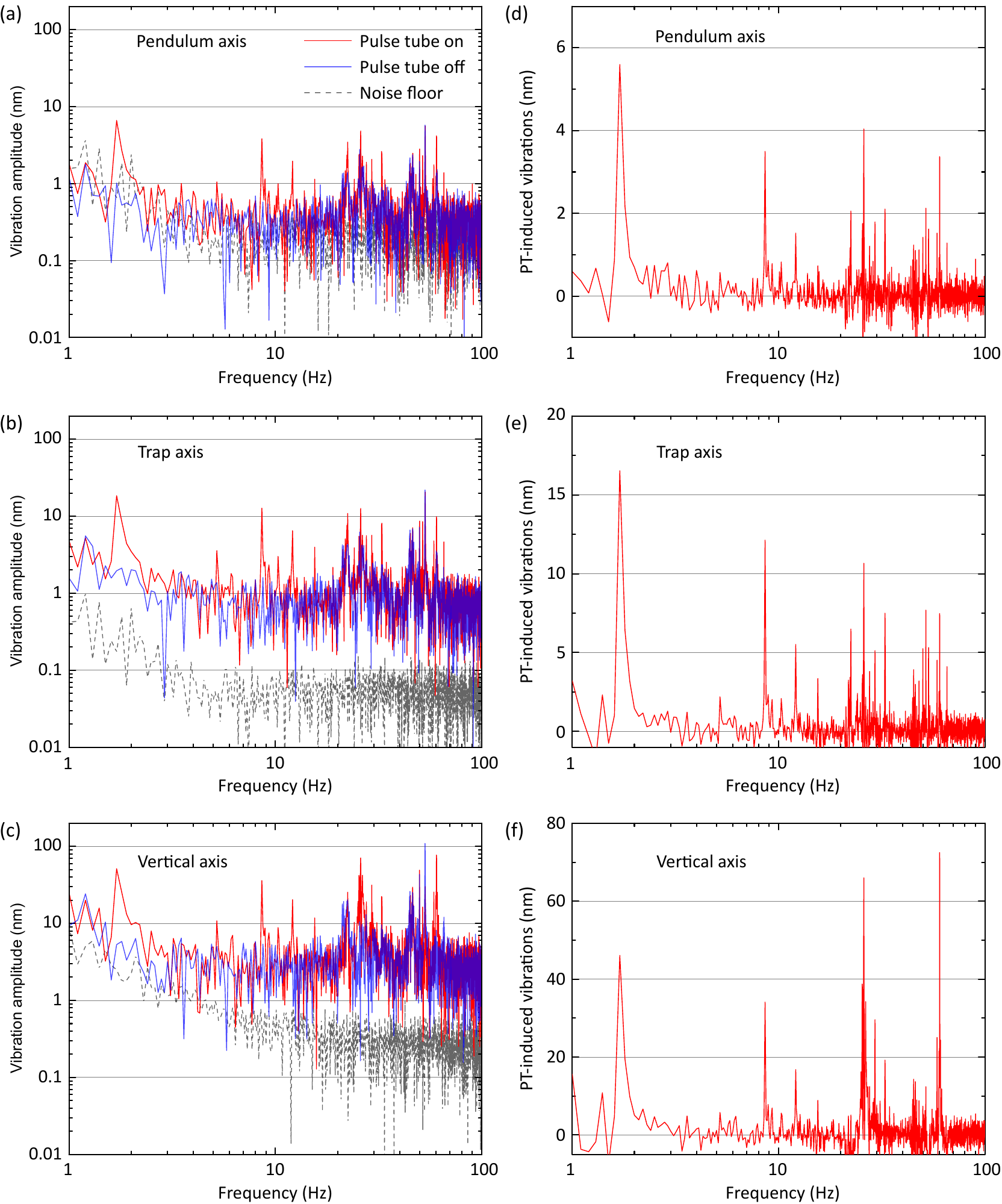}
	\caption{\label{fig:vibration_allaxes}Left column: Vibration amplitude spectra along three orthogonal axes [(a): pendulum axis, (b): trap axis, (c): vertical axis]. The data were taken with the pulse-tube cryocooler on (red) and off (blue) in order to determine its contribution to the vibration levels. Also displayed is the measurement noise floor (dashed), which indicates the sensitivity of the applied method. Right column: Pulse tube-induced vibration amplitudes along the axes to show the effect of the operation of pulse tube and compressor [(d): pendulum axis, (e): trap axis, (f): vertical axis; note the different y-scales]. The spectra were derived from a discrete Fourier transform of 10\,s-long measurements of the mirror displacements with a sampling frequency of 200\,Hz. This results in a corresponding resolution bandwidth of 0.1\,Hz. Note that the spectra for the trap and vertical axes had to be reconstructed by measurements taken at $60^\circ$ and $75^\circ$. See text for further details.}%
\end{figure*}

The inherent vibrations arising in the pulse tube have an amplitude of $\sim$\,10\,\micro m at the cold stages and appear mostly at its operating frequency (1.7\,Hz) plus its harmonics, mostly odd ones (here up to 11th order). The impact on the ion trap of these vibrations, in addition to the other aforementioned noise sources, must be characterized. Vibrations propagating through the TTU and the vacuum chamber walls can cause unacceptable motion of the ion trap. In precision laser spectroscopy for instance, differential vibrations between the optical table and the ion trap lead to Doppler shifts of the interrogating lasers. For coherent operation, the vibration amplitudes should be smaller than the involved wavelengths of the lasers. 

To evaluate the mechanical stability of our system we measured the vibration spectra in three dimensions using self-heterodyne interferometers, with the reference arm located on the optical table and the signal arm connecting to the second stage close by the ion trap using retro-reflecting mirrors. The interferometers employ acousto-optic modulators (AOMs) in the signal arms for frequency shifting twice and silver-coated mirrors installed on three laser access ports for retro-reflection of the incident beam (see scheme in Fig.\,\ref{fig:vibr_setup}). Due to access constraints we could not use mutually orthogonal ports. Thus, simultaneous measurements had to be taken to reconstruct the 3D vibrations along the principal axes. For each single direction, the interferometric beat note was detected with a fast photodiode and demodulated with a frequency mixer driven by twice the AOM frequency. These signals passed proportional-integral (PI) loop filters and were fed to the (phase or frequency) modulation input of the signal generators driving the AOMs, thereby stabilizing the laser phases at the mirror positions inside the vacuum chamber. Depending on the type of feedback provided by the signal generators, displacement amplitudes (phase modulation) or velocities (frequency modulation) were measured. Calibration of the modulation inputs of the signal generators returned absolute displacements. The outputs of the loop filter signals were sent to a fast-Fourier transform device, recording the spectrum of the feedback signals necessary to close the loop, or to an oscilloscope. Fig.\,\ref{fig:timetrace} shows such time traces where the signal, recorded as a voltage, was already converted into displacement amplitudes. As there were only two signal generators with modulation inputs available for the measurement, vibrations on the third axis were measured in an open-loop configuration by observing fluctuations in the error signal after demodulation. The phase of the signal generator output for this dimension was tuned to ensure operation in the linear range of the mixer output.

Vibrations along the horizontal pendulum axis could be directly sampled under an angle of $0^\circ$. Vibrations along the other two principal axes, namely the ion trap axis (also horizontal) and the vertical axis, could be estimated by reconstructing the time traces by vector addition. The measurements for these two axes had to be taken at angles of $60^\circ$ with respect to the trap axis in the horizontal plane and at $75^\circ$ with respect to the vertical axis in the vertical plane through the trap axis, respectively. The subsequent reconstructions involve divisions by $\cos(60^\circ)\approx0.5$ and $\cos(75^\circ)\approx0.26$, also increasing their absolute errors. Consequently, the vertical axis is the least accurate one, however, it is also the least important one since spectroscopy lasers enter the trap in the horizontal plane. In Fig.\,\ref{fig:vibration_allaxes} the vibration amplitude spectra along the three principal axes are displayed, obtained by a discrete Fourier transform of the reconstructed time traces. The left hand column shows the vibrations with the pulse tube switched on and off as well as the measurement noise floor, recorded by blocking the photodiode in Fig. \ref{fig:vibr_setup}. Additionally, the pulse tube-induced vibration spectra are shown in the right hand column, i.\,e. the difference of the spectra with the pulse tube switched on and off. The individual horizontal vibration amplitudes are well below 20\,nm, with the dominating peak at the pulse-tube repetition rate of 1.7\,Hz and $\sim$10\,nm at the fifth harmonic (8.5\,Hz). These oscillations lead to a peak first order Doppler shift on the order of $10^{-16}$. Most of the vibrations at 20 to 60\,Hz can be attributed to resonances of the floor or optical table excited by the compressor. In spectroscopic measurements, vibrations lead to line broadening if their period is shorter than the interrogation time, as it is the case for narrow transitions. Second-order Doppler shifts induced by the measured vibration velocities are negligible.

These results are significantly better than those reported for Gifford-McMahon cryocoolers equipped with helium-gas exchange cells at 4.5\,K operation\cite{pagano_cryogenic_2019}. 

\section{Conclusion}\label{sec:C}
We have introduced a novel closed-cycle cryostat capable of suppressing cryocooler vibrations by about three orders of magnitude to a level of 10\,nm. It provides $< 5$\,K temperatures, an XHV pressure in the $10^{-15}$-mbar range, and 1\,W of cooling power. Installation of noisy components in an adjacent room and several stages of vibration suppression are applied. We installed a radio-frequency ion trap on a pneumatically floating optical table, reproducing its position after a full thermal cycle to room temperature and back to below 5\,K with respect to several aligned laser beams to better than a few \micro m. These conditions are needed for high precision experiments and optical clock operation with HCIs \cite{kozlov_highly_2018}. Our system has stably operated over months, requires little maintenance, and offers convenient vacuum access from the top. To our knowledge, our vibration suppression surpasses all other reported closed-cycle systems with comparable cooling power. The concept can be adapted to other geometries, for example with a vertical pendulum. Further vibration suppression could be achieved by separating the rotary valve unit from the pulse-tube cryocooler, decoupling the pulse tube from the first cross with vertical bellows, employing a cryocooler with lower inherent vibrations, and by mounting segment I and II onto air springs. We are keen to share our design with other research groups.

\begin{acknowledgments}		
We gratefully acknowledge the MPIK engineering design office headed by Frank Müller, the mechanical workshops of MPIK under the direction of Thorsten Spranz, and of PTB headed by Frank Löffler for their expertise and fabrication of numerous intricate parts. We thank the MPIK mechanical apprenticeship workshop, led by Stefan Flicker, where a major number of parts were made and a significant amount of development work took place. We thank in particular Florian Säubert for devising sophisticated procedures for the manufacturing of complex parts. Additionally we appreciated the help by Stephan Metschke, Christian Kaiser, and Alexander Ruhz. We thank Julian Glässel, Michael Drewsen, Timko Dubielzig, and Matthias Brandl for helpful discussions. Financial support was provided by the Max-Planck-Gesellschaft and the Physikalisch-Technische Bundesanstalt. SAK acknowledges support by the Alexander von Humboldt Foundation. We acknowledge support from the Deutsche Forschungsgemeinschaft through SCHM2678/5-1 and the Collaborative Research Centre “SFB 1225 (ISOQUANT)”.
\end{acknowledgments}

\end{document}